\documentclass[aps,pra,twocolumn,groupedaddress,showpacs,superscriptaddress,scrartcl]{revtex4-1}
\usepackage{graphicx}
\usepackage{comment,amssymb,amsmath,graphicx,epstopdf,hyperref,amsthm,dsfont,enumerate,bm,color}
\begin{document}
\interfootnotelinepenalty=10000

\title{Flexible quantum tokens in spacetime}

\author{Adrian Kent} 
\email{apak@damtp.cam.ac.uk} 
\affiliation{Centre
  for Quantum Information and Foundations, DAMTP, Centre for
  Mathematical Sciences, University of Cambridge, Wilberforce Road,
  Cambridge, CB3 0WA, U.K.}  
\affiliation{Perimeter Institute for
  Theoretical Physics, 31 Caroline Street North, Waterloo, ON N2L 2Y5,
  Canada}

\author{Dami\'an Pital\'ua-Garc\'ia}
\email{D.Pitalua-Garcia@damtp.cam.ac.uk} 
\affiliation{Centre for
  Quantum Information and Foundations, DAMTP, Centre for Mathematical
  Sciences, University of Cambridge, Wilberforce Road, Cambridge, CB3
  0WA, U.K.}

\date{\today}

\begin{abstract}
S-money \cite{kent2019s} schemes define virtual tokens
designed for networks with relativistic or 
other trusted signalling constraints.   
The tokens allow near-instant verification and guarantee
unforgeability without requiring quantum state storage.  
We present refined two stage S-money schemes.   The first
stage, which may involve quantum information exchange, generates private user token data.
In the second stage, which need only involve classical communications,
users determine the valid presentation point,
without revealing it to the issuer.  
This refinement allows the user to determine the presentation point
anywhere in the causal past of all valid presentation
points.   It also allows flexible transfer of tokens
among users without compromising user privacy.
\end{abstract}

\maketitle

\section{Introduction}

Money and other type of tokens allow a user to access a resource,
while guaranteeing to the issuer that the tokens cannot be forged.
Quantum money and quantum token 
schemes \cite{wiesner1983conjugate, gavinsky2012quantum, pastawski2012unforgeable, GK15} can theoretically guarantee
unforgeability with unconditional security.  However, standard quantum
money and quantum token schemes are impractical with current
technology because they require long-term quantum state storage. 
Quantum memories cannot currently store quantum states
reliably for more than a fraction of a second. 
Although longer term quantum memories will presumably
be developed in future, it is unclear whether (or when)   
quantum memory technology will ever be competitive in
price and practicality with classical alternatives. 

S-money \footnote{Depending on context, `S' may stand
for `summonable' or `super'.} schemes 
\cite{kent2019s, Kentquantumtokens, Kentfuturepositioncommitment, Kentmethodandsystem} offer alternatives to quantum money that do not require
quantum memory and can respond more flexibly to generalized summoning
tasks \cite{kent2019s}.  
However,  
many other theoretical and practical questions about the
relative advantages, implementability and resource costs of S-money
and quantum money token schemes remain to be addressed.   
In our view, both technologies are potentially promising and 
merit further refinement and development.   

In this paper we describe refinements of S-money schemes that 
improve their flexibility and transferability.     
We focus on a single round implementation of a simple quantum S-money token scheme in order to 
illustrate the essential ideas; it is straightforward to 
extend these to multiple rounds, to other types of S-money token, to schemes involving
subtokens and to more complex scenarios.
We also simplify the discussion by focusing on 
an S-money scheme that guarantees unforgeability and
future privacy for the user, without the extra cryptographic
layer required to ensure past privacy; again, the extension
is straightforward.  

\section{Flexible S-money}
We follow the standard scenario for S-money \cite{kent2019s}.
The issuer (Bob the bank) consists of a network of local agents
based at secure sites, with authenticated and secure communication links, working in collaboration and
with mutual trust.   Similarly, the user (Alice the acquirer) consists of a network of
local agents based at secure sites, with authenticated and secure communication links,
working in collaboration and with mutual trust.  
The user and issuer agree to implement the scheme but 
do not trust one another to behave honestly.   
The user's secure sites do not overlap with the issuer's. The user and issuer agree on a reference frame, where spacetime coordinates are defined. The user and issuer agree in advance on a set of presentation
points $Q_i$, where the user can present an S-money token to the issuer in order to
access a resource.   
The scheme begins at some start point $P$ with $P \prec Q_i$ (that is, $P$
is in the causal past of $Q_i$) for all $i$.   
To be precise, both $P$ and the $Q_i$ actually represent localized
spacetime regions containing separate secure sites for the user and
issuer, with a time window allowing prescribed communications between
these sites.   Because the spatial and/or temporal separations 
between these regions are large (in the relevant rest frame)
compared to those of the regions, we can think of them as effectively
points on a spacetime network.    
The token scheme is initiated by classical and/or quantum information
communications between user and issuer agents at $P$. 
To consider transferability of tokens we refer to two or more users
(Alice${}_1$, Alice${}_2$, $\ldots$).   In this case each user has 
their own separate secure site at each relevant network point.

\subsection{Bit-string coordination and S-money schemes}
These schemes are based on bit-string coordination \cite{kent2019s} protocols between
the user and issuer.
Bit-string coordination is an intrinsically relativistic task
in which the user inputs, or \emph{commits}, a string $b$ at some point in spacetime, 
and later \emph{unveils} by giving classical data to the issuer to show that her input was
$b$. A secure implementation of bit-string coordination satisfies two security properties: binding and hiding. We say a protocol is \emph{binding} if it guarantees to the issuer
that if the user successfully proves having input $b$ then
she cannot also prove (either at the same spacetime point or elsewhere)
having input a different string $b'$. We say a protocol is \emph{hiding} if it guarantees to the user that the issuer cannot know $b$ until it is unveiled by the user. Bit-string coordination is similar to but weaker than bit-string
commitment, which also needs to guarantee to the issuer
that the user committed to her input at some specified start
point. In this paper we apply the language of bit-string commitment to bit-string coordination, as specified above.

In the S-money schemes \cite{kent2019s, Kentquantumtokens} we consider
here, the user defines their bit string by choices of measurements on
quantum states sent by the issuer. The schemes and our refinements are
secure regardless of the user's technological limitations. However, to
simplify the discussion, we will assume the user is technologically
limited and has no quantum state storage, which effectively forces her
to carry out these measurements on receipt.  Given this limitation,
the bit string coordination protocol effectively becomes a bit string
commitment protocol, with the commitment being made when the states
are measured.  Specifically, the commitment phase for each bit in the
string is that of the BB84 bit commitment protocol of
Ref. \cite{kent2012unconditionally}, whose security was analysed in
Refs. \cite{kent2012unconditionally, croke2012security,
  kaniewski2013secure, lunghi2013experimental}.  This protocol was
implemented experimentally \cite{lunghi2013experimental} with a
modification involving precommitment to a random string in order to
allow practical implementation without requiring long distance quantum
communication. This technique was used also used in practical
protocols for spacetime-constrained oblivious transfer
\cite{PGK18,PG19}. We use variations on this idea here.

\subsubsection{Example of a bit coordination scheme}
\label{Exex}
We present a simple bit coordination scheme given in
Ref. \cite{Kentquantumtokens}, which is based on the bit commitment
protocol of Ref. \cite{kent2012unconditionally}, to 
illustrate why bit-string coordination and bit-string commitment are
different tasks.

Let
$\{\lvert 0\rangle,\lvert 1\rangle,\lvert +\rangle,\lvert -\rangle\}$
be the set of Bennett-Brassard 1984 (BB84) states, where
$\lvert \pm\rangle=\frac{1}{\sqrt{2}}\bigl(\lvert 0\rangle\pm\lvert
1\rangle\bigr)$
and where the states $\lvert 0\rangle$ and $\lvert 1\rangle$ are
orthonormal. We define the following qubit states
\begin{equation}
\label{BB84}
\lvert\phi_{0}^{0}\rangle=\lvert 0\rangle,\quad 
\lvert\phi_{1}^{0}\rangle=\lvert 1\rangle,\quad\lvert \phi_{0}^{1}\rangle=\lvert +\rangle,\quad\lvert \phi_{1}^{1}\rangle=\lvert -\rangle.
\end{equation}
Let $n\in\mathbb{N}$ be a security parameter. For strings
$\bold{r},\bold{s}\in\{0,1\}^{n}$, let their respective bit entries be
denoted by $r_l,s_l\in\{0,1\}$, for $l\in[n]$, where we define
$[n]= \{ 1, 2, \ldots , n \}$.
We define the state
\begin{equation}
\label{new}
\lvert \Psi_{\bold{r},\bold{s}}\rangle=\bigotimes_{l\in[n]} \Big\lvert \phi_{r_l}^{s_l}\Big\rangle,
\end{equation}
for $\bold{r},\bold{s}\in\{0,1\}^{n}$. 

The issuer generates a state
$\lvert\Psi\rangle=\lvert \Psi_{\bold{r},\bold{s}}\rangle$ of $n$
qubits, with $\bold{r}$ and $\bold{s}$ chosen randomly from
$\{0,1\}^{n}$, and sends $\lvert\Psi\rangle$ to the user at a
space-time
point $P$. The issuer
sends the labels $l$ of the transmitted qubits, for $l\in[n]$. The
user commits to the bit $b=0$ by measuring all received qubits in the
computational basis ($\{\lvert 0\rangle,\lvert 1\rangle\}$), or to the
bit $b=1$ by measuring all received qubits in the Hadamard basis
($\{\lvert +\rangle,\lvert -\rangle\}$). Let $y_l$ be the bit
measurement outcome corresponding to the qubit labeled by $l$, for
$l\in[n]$, and let $\bold{y}=(y_1,\ldots,y_n)$.  The user unveils 
at some point $Q \succ P$ by giving the bit $b$ and the string of outcomes $\bold{y}$ to the
issuer. In the ideal case that errors are not tolerated, the issuer
accepts $b$ as valid if $y_l=r_l$ for all $l\in[n]$ satisfying
$s_l=b$. This means for example that if the user unveils $b=0$, and he
is being honest, he measured all qubits in the computational basis;
hence, the user obtains the correct outcomes for all qubits prepared
by the issuer in the computational basis. In a realistic case, the
issuer accepts $b$ as valid if it holds that $y_l\neq r_l$ for no more
than $\gamma \lvert \Omega_b\rvert$ qubits with labels $l\in\Omega_b$,
where $\Omega_b=\{l\in[n]\vert s_l=b\}$, for a predetermined small
enough $\gamma\in(0,\frac{1}{2})$.

This bit coordination protocol is hiding because the user does not
give any information to the issuer about her chosen bit $b$ before
she unveils.  It is binding because the user cannot, at any two points
$Q, Q'$, give the
issuer two $n-$bit strings $\bold{y}=(y_1,\ldots,y_n)$ and
$\bold{z}=(z_1,\ldots,z_n)$, corresponding respectively to unveiling
$b=0$ and $b=1$, that satisfy both conditions 1) $y_l\neq r_l$ for no
more than $\gamma \lvert \Omega_0\rvert$ qubits with labels
$l\in\Omega_0$, and 2) $z_l\neq r_l$ for no more than
$\gamma \lvert \Omega_1\rvert$ qubits with labels $l\in\Omega_1$, for
a predetermined small enough $\gamma\in(0,\frac{1}{2})$. This property follows
from the security analyses given in
Refs. \cite{kent2012unconditionally, croke2012security,
  kaniewski2013secure, lunghi2013experimental,KLPGR19}.
  
However, this does not define
an unconditionally secure bit commitment protocol, since the issuer 
receives no useful guarantee on the spacetime region where the user chose
$b$.  For 
example, a technologically unlimited user could store all
received states in a perfect quantum memory, send them to $Q$, and
measure them there in the relevant basis just before unveiling $b$.   
The issuer is guaranteed that the bit was chosen at some point $Q'
\preceq Q$, but not that it was chosen at $P$ or any other specified point
$Q' \prec Q$.

\subsection{Limitations of previous S-money schemes}

\subsubsection{Initiation time} 

The bit coordination protocol just discussed can be
repeated $N$ times, giving an $N$ bit string coordination protocol
\cite{kent2019s}.   This defines a simple S-money token protocol
\cite{kent2019s, Kentquantumtokens}, 
in which the bit string labels the point $Q_i$ (for $i \in \{0,1\}^N$)
at which the token may validly be presented.  
The S-money scheme is initiated by the issuer
sending a string of quantum states to the user at
the start point.   In the ideal version, these states are
pure qubits drawn randomly from the four BB84 states, sent
along a lossless channel, and the user carries out perfect
BB84 basis measurements on each state.   
In practical embodiments with photonic systems, this quantum communication step
involves generating a sequence of weak photon pulses, with some errors
in the polarizations chosen, some losses, and some measurement errors.
Depending on the experimental parameters, a reasonable level of
security may require the transmission 
and measurement of a large number of photon pulses.
With current technology, this could take at least a few seconds. 
In an S-money scheme, this would require the user to choose her presentation point
$Q_b$ at least a few seconds in the past of $Q_b$. 

This is a significant and undesirable limitation, 
since S-money schemes are intended for application
in relativistic scenarios where time is critical 
and where decisions about when and where to present the
S-money are ideally made as flexibly as possible (and
in particular as late as possible) on the basis of
incoming information received at various spacetime points.

Quantum technology is developing rapidly, and it may be possible
to generate, send and measure enough quantum states to initiate secure tokens within very short time intervals in 
future.   Nonetheless, we think it likely that the alternatives we
discuss below, which require only classical communications at the 
critical decision point(s), will continue to be faster and cheaper.
Our discussion and illustrations rely only on this assumption,
rather than on specific quantitative estimates.   

\subsubsection{Transferability and Transfer time}

A related limitation of the S-money schemes \cite{kent2019s,
Kentquantumtokens}, as presented, is that the process of transferring
tokens either lacks flexibility or requires a delay to initiate a new
token.   

Suppose that $B$ issues a token to $A_1$, by sending a string of 
quantum states within a (suitably extended) space-time region $P$.  We assume that $A_1$ has no long-term quantum memory, since
this is the underlying motivation for the schemes.  
Generating a token thus requires $A_1$ to measure the BB84 states sent
by $B$ as soon as she receives them, in bases that label her
chosen presentation point $Q_b$. 

Suppose now that $A_1$ later wishes to transfer the token to $A_2$ at 
some point $Q$, where $P \prec Q \prec Q_b$.   Two options are 
considered in Ref. \cite{kent2019s}.  

1. $A_1$ may give $A_2$ the rights to a token which
is valid only at her already determined presentation point, $Q_b$.    
She can do this by giving $A_2$ at $Q$ her measurement
outcomes and the number $b$, along with a digitally signed
classical message stating that $A_1$ relinquishes her token rights
and $A_2$ is now the owner.   $A_2$ can send this signed message to 
a local agent of $B$ at or near $Q$, who informs all agents of
$B$ in his causal future.  These agents will then no longer accept
the token from any agent of $A_1$, but will accept it from $A_2$. 

2. $A_1$ may give $A_2$ the rights to the token at $Q$ in a way that allows
$A_2$ to choose a new valid presentation point, $Q_{b'} \succ Q$.  
She can do this by giving $A_2$ a digitally signed message stating
that she relinquishes her token rights and that the token is no
longer valid.  $A_2$ sends this to a local agent of $B$, as above.   
In this case, the local agents of $B$ and $A_2$ need to initiate
a replacement token.  This requires $B$'s local agent to send
a string of quantum states to $A_2$'s local agent, who measures
them, following the protocol above.  

The first option has the advantage that transfers require only
classical communications, and so can be completed relatively quickly.
However, it is inflexible; $A_2$ is restricted to presenting the token
at the point $Q_b$ previously chosen by $A_1$.

The second has the advantage of flexibility; $A_2$ may choose a new
presentation point, which may be any valid presentation point within
the causal future of the transfer point $Q$.   However, it requires
a new round of quantum communications, with the associated delay.   

\subsection{Flexible S-money schemes}

In this paper, we propose an implementation of 
S-money that has none of the above disadvantages.   
The scenario we envisage requires that users may realise they may potentially want to
acquire and use S-money well in advance of actually doing so.
This seems generally realistic.   For example, traders who plan
high value high speed trades on the global financial network 
expect to be registered, authenticated, licensed, to set up infrastructure,
and so on, before they begin trading: they do not expect to 
be able to show up on the network unannounced and uncredentialled
and instantly trade.  

We thus propose that such users go through information exchanges with
the bank, in which the bank sends strings of quantum states which the
user measures, in some starting region $S$, well before they might
acquire or use S-money.
These exchanges define bit string coordination protocols which, if the
user is unable to store quantum states, effectively commit 
the user to long random bit strings $x$.  
To simplify the discussion, we will assume the users have this
technological limitation.  However, this is not required for the
security of the schemes, which (like the original schemes discussed
above) require
only secure bit string coordination, not bit string commitment.  
The bank's data received from each
user's protocol are shared with all the bank's local agents, but kept private 
from other users.       

To acquire an S-money token, a user $A$ then needs only to
communicate classically with the bank at some later point $P_A$.
They first agree a set of valid presentation points $\{ Q_i : i \in S \}$.
Here $S$ satisfies $2^{M-1} < | S | \leq 2^M$, each $Q_i \succeq
P_A$, and they agree a standard convention for the labeling of the $Q_i$
by bit strings $i$.   
They also agree which previously unused
length $M$ segment $x$ of the user's committed bit string will 
be associated with this token, by labeling the relevant bits.   
(The user keeps the value of $x$ secret here.) 
They may also at this point agree the price to be paid (e.g. in some standard currency)
for the S-money token, and perhaps process the payment.
The bank's local agent sends all the information received to 
the bank's agents at each $Q_i$.  

The user may then later decide where the token will be valid,
and commit to this decision by choosing $b\in S$, at any network point $P_D$ such that 
$P_A \preceq P_D \preceq Q_i$ for all $i \in S$. 
The user's local agent at $P_D$ commits this to the bank's local agent
by sending him the string $m = x \oplus b$, and the bank's local agent
sends this to the bank's agents at each $Q_i$, where `$\oplus$' denotes sum modulo two. 

Finally, at $Q_b$, the user's local agent unveils the commitment 
to $x$ by sending their commitment data to the bank's local agent.
The bank's local agent verifies the commitment and that 
$m \oplus x = b$; if these tests are passed, the S-money token
is validated at $Q_b$, and the user is given whatever resources
were agreed there.

\subsubsection{Example}

In the token scheme presented in Ref. \cite{Kentquantumtokens} the
issuer transmits a quantum state $\lvert \Psi\rangle$ to the user,
chosen from a predetermined set. At reception of $\lvert \Psi\rangle$,
the user chooses the number $b$ labeling the presentation point $Q_b$
by applying on $\lvert \Psi\rangle$ a quantum measurement ${\rm M}_b$ that
belongs to a predetermined set, obtaining a classical measurement
outcome $\bold{y}$.\footnote{In practical implementations with
current technology, the
transmitted state will be a sequence of (approximate) qubits, and
the measurement a sequence of individual measurements on these.   
We represent these as a single state and single collective measurement
to simplify the notation.}
The user presents the token at $Q_b$ by presenting
the classical measurement outcome $\bold{y}$ at $Q_b$. The issuer
validates the token at $Q_b$ if he verifies that the data $\bold{y}$
received from the user at $Q_b$ corresponds to a valid measurement outcome 
of  the quantum measurement M$_b$ applied
on $\lvert \Psi\rangle$.

The following example to implement the previous token scheme
  is given in Ref. \cite{Kentquantumtokens}. Let the presentation
  points $Q_i$ be labeled by strings $i=(i_1,\ldots,i_M)\in\{0,1\}^M$,
  for an integer $M$ predetermined by the user and issuer.  Let
  $b=(b_1,\ldots,b_M)\in\{0,1\}^M$ be the string that labels the
  user's chosen presentation point $Q_b$. The user and issuer engage
  in $M$ parallel protocols of the bit coordination scheme presented
  in section \ref{Exex}. The $k$th bit coordination protocol defines
  the bit entry $b_k$ of the user's string $b$, for $k\in[M]$. Thus,
  more precisely, the issuer generates a state
  $\lvert\Psi\rangle=\lvert \Psi_{\bold{r},\bold{s}}\rangle$ of $nM$
  qubits given by
\begin{equation}
\label{new2}
\lvert \Psi_{\bold{r},\bold{s}}\rangle=\bigotimes_{(k,l)\in[M]\times[n]} \Big\lvert \phi_{r_l^k}^{s_l^k}\Big\rangle,
\end{equation}
where $\bold{r}$ and $\bold{s}$ are chosen randomly from
$\{0,1\}^{Mn}$ by the issuer, and where $r_l^k$ and $s_l^k$ denote the
respective bit entries with labels $(k,l)$ of the strings $\bold{r}$
and $\bold{s}$, for $k\in[M]$ and $l\in[n]$. The issuer sends to the
user the state $\lvert\Psi\rangle$ and the labels $(k,l)$ of the
transmitted qubits, for $k\in[M]$ and $l\in[n]$. The user's
measurement M$_b$ consist in measuring the received qubit with label
$(k,l)$ in the computational basis
($\{\lvert 0\rangle,\lvert 1\rangle\}$) if $b_k=0$ or in the Hadamard
basis ($\{\lvert +\rangle,\lvert -\rangle\}$) if $b_k=1$,
respectively, for $l\in[n]$ and for $k\in[M]$.

 In our flexible S-money scheme, at reception of the token from the
 issuer, the user applies the quantum measurement M$_x$ on the quantum
 state $\lvert \Psi\rangle$ received from the issuer, where $x$ is a
 random $M-$bit string chosen by the user. The user obtains a
 classical measurement outcome $\bold{y}$. Then, the user chooses the
 $M-$ bit string $b$ labeling her presentation point $Q_b$ and sends
 $m=x\oplus b$ to the issuer at a point in the causal
   future of the point where she completed the
 measurement M$_x$ and within the intersection of the causal pasts of
 all the presentation points. The user gives the token $\bold{y}$ to
 the issuer at $Q_b$. The issuer validates the token if he verifies
 that $\bold{y}$ is a statistically plausible measurement outcome of
 the quantum measurement M$_{m \oplus b}$ applied to 
 $\lvert \Psi\rangle$. This is illustrated in Fig. \ref{fig1} below.

\begin{figure}
\includegraphics{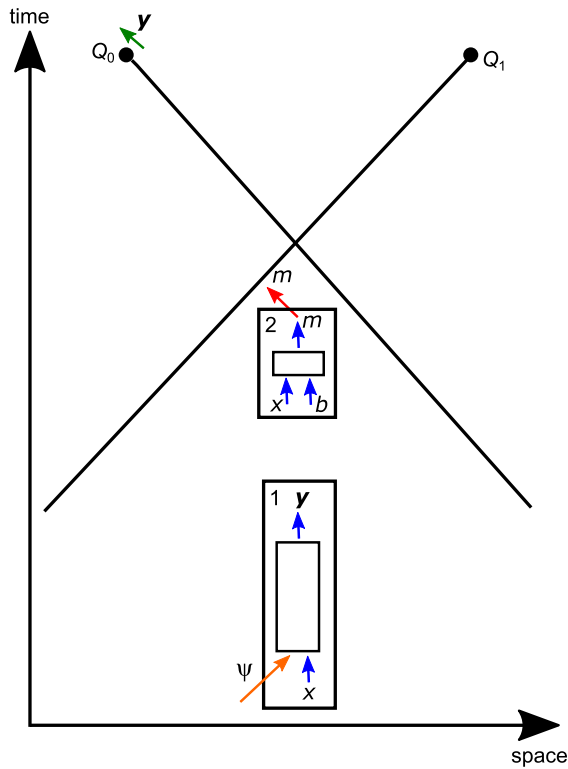}
\caption{\label{fig1} Our flexible S-money scheme
    described in the main text is illustrated in $1+1$-dimensional
    Minkowski spacetime for the case of two presentation points $Q_0$
    and $Q_1$ (black dots). We use units in which the speed of light is unity. The black lines denote light rays. Bob's
    transmission of the quantum state $\lvert \Psi\rangle$ to Alice is
    illustrated with the large orange diagonal arrow. The measurement
    on $\lvert \Psi\rangle$ by Alice is represented by the rectangle
    labeled `1'. The generation of the classical message $m=x\oplus b$
    by Alice is represented by the square labeled `2'. Alice's
    transmission of $m$ to Bob is illustrated with the red diagonal
    arrow. The transmission of the token consists in Alice giving her
    measurement outcome $\bold{y}$ to Bob at $Q_b$ (green diagonal
    arrow).  The case $b=0$ is illustrated. The box labeled
    `1' illustrates Alice's generation of the random string $x$ and
    the application of the measurement M$_x$ on the received quantum
    state $\lvert \Psi\rangle$, which outputs a classical measurement
    outcome $\bold{y}$. The box labeled `2' illustrates Alice's
    computation of the message $m=x\oplus b$, where $b$ labels her
    chosen presentation point $Q_b$. Alice's inputs, $x$ and $b$, and
    outputs, $m$ and $\bold{y}$, are represented by the blue vertical
    arrows.}
\end{figure}

\subsection{Security of flexible S-money schemes}

\subsubsection{Unforgeability}
\label{sec3.2}

A refined S-money scheme based on a bit-string coordination protocol
that is binding satisfies the property of unforgeability, according to
which the user cannot have two or more tokens validated at two or more
presentation points, as we discuss. In a cheating strategy by the user
trying to present tokens at two different presentation points $Q_{i_1}$
and $Q_{i_2}$, the user must be able to present unveiling data that
correspond to a valid commitment to a number $x_1$, satisfying
$m \oplus x_1=i_1$, and also to a number $x_2$, satisfying
$m \oplus x_2=i_2$. Since $i_1 \neq i_2$, we have that $x_1\neq x_2$.
However, if the bit-string coordination protocol is binding then, by
definition, if the user presents unveiling data that corresponds to a
valid commitment to a number $x_1$ then the user cannot also present
unveiling data that corresponds to a valid commitment to a number
$x_2\neq x_1$. Thus, the property of unforgeability follows.

\subsubsection{Future privacy for the user}
\label{sec3.3}

A refined S-money scheme based on a bit-string coordination protocol
that is hiding and where $x$ is chosen randomly and securely by the
user satisfies the property of future privacy for the user, according
to which the issuer cannot obtain any information about the presentation point $Q_b$ chosen by
the user before the token is presented by the user, as we discuss. In
a cheating strategy by the issuer trying to obtain some information
about the number $b$ labeling the presentation point $Q_b$ chosen by
the user, the issuer may try to learn some information about the
number $x$ to which the user commits at the reception of the token in
the bit-string coordination protocol and use this obtained information
together with the message $m=x\oplus b$ received from the user to
obtain some information about $b$. However, if we assume that the
bit-string coordination protocol is hiding then, by definition, the
issuer cannot know the value of $x$ before it is unveiled by the
user. Therefore, assuming that $x$ is chosen randomly and securely by
the user, the issuer cannot obtain any information about $b$ from the
message $m=x\oplus b$, before $x$ is unveiled by the user. Thus,
future privacy for the user follows.

\section{Transferring S-money tokens}
\label{sec4}

Our refinement allows S-money tokens to be traded and transferred
between users quite simply and flexibly. 
Suppose the first user $A_1$ has acquired a token at point $P$, 
associated with her committed bit string $x_1$, and
now wishes to transfer it to a second user $A_2$ at point $P_T$, 
where $P \preceq P_T \preceq Q_i$ for all $i$.   
In the simplest version, $A_1$'s local agent at $P_T$ simply
gives $A_2$'s local agent all the data defining the labels
of the presentation points in $S$, and an authenticated digitally
signed message transferring her rights in the token. $A_2$ verifies this message and registers the transfer 
with the bank's local agent, agreeing to associate to the token a labeled segment
$x_2$ of $A_2$'s precommitted random string.  
The local agents of $A_1$ and $A_2$ may also at this point agree the price to be paid (e.g. in some standard currency)
for the S-money token transfer, and perhaps process the payment.
The bank's local agent notifies all agents at points $Q_i$
of the transfer and of the labeled string segment.   

Now $A_2$ may decide the valid presentation point $Q_b$
at any point $P_D$ such that $P_T \preceq P_D \preceq Q_i$ for all $i$, and commit to this
by sending the string $m_2 =  x_2 \oplus b$ to Bob. 
She may present the token at $Q_b$ by sending all the commitment
data for  $x_2$ 
(which comes from her own commitments) to Bob's local agent there.
Bob's local agent validates both commitments, and accepts the
token if the commitments are validly unveiled and
$m_2 \oplus x_2=b$. 

Alternatively, instead of deciding a presentation point, 
$A_2$ may transfer the token to $A_3$ at some transfer 
point $P_{T'} \succeq P_T$ by sending all the relevant data
and her own authenticated digitally signed transfer message,
and so on. 

Each user $A_i$ is guaranteed future privacy regarding
their token presentation point, since 
when they commit to the valid presentation point $Q_b$ at
some point $P_D$ they send a string $m_i =  x_i \oplus b$, 
where $x_i$ is a committed private random string. 

\begin{figure}
\includegraphics{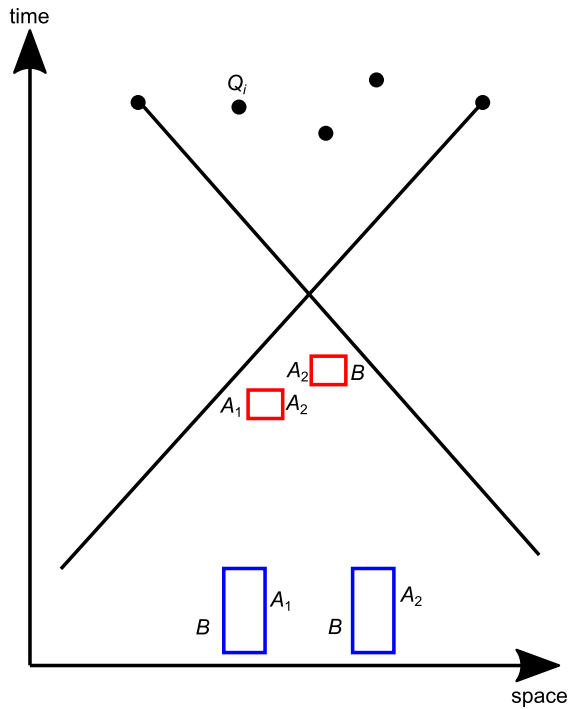}
\caption{ \label{flexiblefig3} Schematic illustration  in $1+1$-dimensional
    Minkowski spacetime of
    transfer of an S-money token in our flexible scheme.  We use units in which the speed of light is unity. The black lines denote light rays. The bank sends quantum states to potential users $A_1 , A_2 , \ldots$
to initiate bit string coordination protocols, which effectively
commit
these users to random classical bit strings $x_1 , x_2 , \ldots$ These protocols (represented by the large blue boxes) may be relatively lengthy, but can be carried out
well before tokens are generated or transferred.   
If $A_1$ initiates a token protocol (not illustrated), she may later
wish to transfer the token to another user $A_2$.   This requires only
classical communications (represented by the small red boxes) between local agents of $A_1$ and $A_2$ and
between local agents of $A_2$ and $B$.  Such communications can be
completed relatively quickly, leaving $A_2$ in control of the token
and free to make her own choice of a presentation point.}
\end{figure}

The key advantage is summarised schematically in
Fig. \ref{flexiblefig3}. 

\section{Delaying the choice of presentation point}
\label{sec6}

This technique extends to allow the user to make a sequence
of decisions and commitments at separate points in spacetime that collectively
decide the presentation point, while still guaranteeing to the issuer 
that there can be no more than one valid presentation point.

Let $P_{D_i}$ for $i=1, \ldots , N$ be a set of decision points,
with at least one of them, $P_{D_1}$, in the causal past of
all valid presentation points. Let $\{Q_i : i\in S\}$ be the set of presentation points. The user may decide and commit at $P_{D_1}$ to restrict the
valid presentation point to lie in a subset
$\{ Q_i : i \in S_1 \}$, sending the issuer 
a description of the subset $S_1$ committed using data from their precommitted
random string $x$.   Let $P_{D_2}$ lie in the causal past 
of all $Q_i$ with $i\in S_1$. The user may decide and commit at $P_{D_2}$ to restrict the valid presentation point to lie in a subset
$\{ Q_i : i \in S_2 \}$, sending the issuer 
a description of the subset $S_2 \subset S_1$ committed using further data from their precommitted
random string, and so on, with the valid presentation point finally
decided and committed to at (or before) the final point 
$P_{D_m}$.   

These commitments may be coded efficiently 
if the configuration of the decision points and the
sizes and relations of the relevant subsets are 
suitable and known in advance. 
For example, if it is known in advance
that a binary choice will be made at each successive decision point,
selecting one of two equally sized known subsets, then the 
user may simply commit using successive bits of the string
$x$ that would have been used to commit directly to $b$ in
the unrefined protocol. 

\subsection{Example}

\begin{figure}
\includegraphics{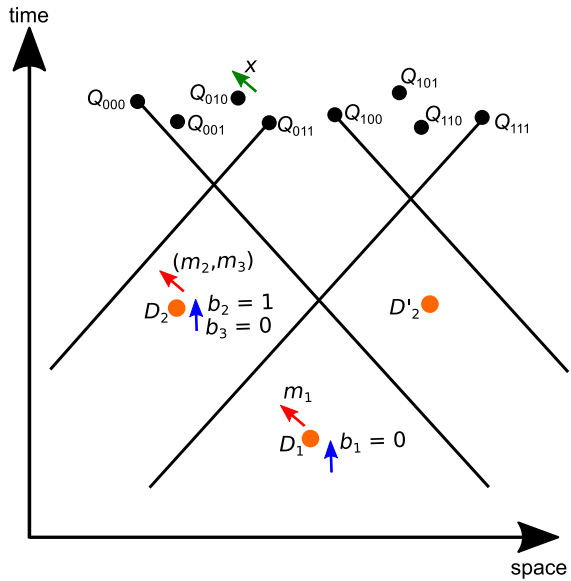}
\caption{\label{fig2} The refinement allowing Alice to delay her
  chosen presentation point described in the main text is illustrated
  in $1+1$-dimensional Minkowski spacetime. We use units in which the speed of light is unity. In
  this example, there are eight presentation points (black dots). The
  black lines denote light rays. Alice has spacetime decision points,
  $D_1$ and $D_2$ (orange dots). $D_1$ is in the causal past of all the
  presentation points. $D_2$ is in the causal past of $Q_{0i_2i_3}$, for all
  $i_2,i_3\in\{0,1\} $. Alice's choices for the bit entries of the
  label $b=(b_1,b_2,b_3)=(0,1,0)$ of her chosen presentation point
  $Q_b$ are illustrated with the blue vertical arrows. The bit
  $m_1=x_1\oplus b_1$ and the two-bit string
  $(m_2,m_3)=(x_2,x_3)\oplus(b_2,b_3)$ that Alice sends Bob at
  $D_1$ and $D_2$, respectively, are indicated by the red diagonal
  arrows. Presentation of the token by Alice to Bob at $Q_b$ consists
  in Alice unveiling her three bit string $x$ to Bob at the
  presentation point $Q_b$, as illustrated with the green diagonal
  arrow.  An alternative decision point, $D'_2$, is also
    shown. Alice may use this if she chooses $b_1=1$ at $D_1$.}
\end{figure}

In Fig. \ref{fig2}, we illustrate a simple example with eight
presentation points $Q_i$, with $i\in S=\{0,1\}^3$, and two decision
points, $N=2$. Let $x,b\in\{0,1\}^3$ and let $i_k$ be the $k$th bit entry of $i$, for $k\in\{1,2,3\}$ and for $i\in\{x,b\}$. $D_1\equiv P_{D_1}$ is a
spacetime point within the intersection of the causal pasts of all the
presentation points, where Alice decides that her presentation point
$Q_b$ will belong to the set
$\{Q_{b_100},Q_{b_101},Q_{b_110},Q_{b_111}\}$, i.e. with
$S_1=\bigl\{\bigr (b_1,i_2,i_3): i_2,i_3\in \{0,1\}\}$.
In this example, we suppose she chooses $b_1=0$. At $D_1$, Alice indicates to Bob that she has chosen
her presentation point to have the bit $b_1$ fixed, without telling
him the chosen value for $b_1$, and sends him the bit
$m_1=x_1\oplus b_1$, where $x_1$ is the first bit of the
previously committed string $x=(x_1,x_2,x_3)$. At the decision point
$D_2\equiv P_{D_2}$, which is in the causal past of all
$Q_i$ with $i\in S_1$, Alice chooses the bits $b_2$ and $b_3$ of her
presentation point $Q_b$, with values $b_2=1$ and $b_3=0$ in this
example, she indicates to Bob that she has chosen her presentation
point and sends the two-bit string
$(m_2, m_3)=(x_2,x_3)\oplus(b_2,b_3)$ to Bob. Alice presents her token
at $Q_b$ by unveiling to Bob's agent at $Q_b$ her commitment to the
string $x$. Bob's agent at $Q_b$ validates Alice's token at $Q_b$,
with $b=(b_1,b_2,b_3)$, if he validates Alice's commitment to $x$ and
if $m_i\oplus x_i =b_i$ for $i=1,2,3$.  

\subsection{Comparison with previous S-money schemes} Previous
  S-money schemes can also allow users to delay the choice of
  presentation point.  For example, Ref. \cite{Kentquantumtokens}
  considers a scheme in which the user may present the token at a
  presentation point $Q_b$ by giving an issuer's agent there her
  string $\bold{y}$ of measurement outcomes, or she may delay the
  presentation of her token by stating this to the issuer's agent at
  $Q_b$. In the latter case, the issuer's agent gives further quantum
  states to the user at $Q_b$, and the user measures the received
  quantum states at $Q_b$ in a basis labeling her chosen presentation
  point $Q_{b'} \succ Q_b$, obtaining a string of measurement outcomes
  $\bold{y}'$. At $Q_{b'}$, the issuer may present the token by giving
  the issuer's agent there the string $(\bold{y},\bold{y}')$, or she
  may choose to further delay the presentation point, and so on.

  This method has the disadvantage that it requires various quantum
  communication exchanges between user's and issuer's agents at
  various spacetime points. This is technologically very demanding
  because the issuer and user's agents must have quantum devices to
  transfer and measure quantum states at all presentation points where
  the token presentation can be delayed. On the other hand, our scheme
  only requires the quantum communication between user and issuer to
  take place once, between a single pair of adjacent laboratories at a
  spacetime region $P$, so that delays of the presentation point
  are needed only to allow time for classical communications between user and
  issuer's agents at spacetime points in the causal future of $P$.


\section{Discussion}
\label{sec7}
In this paper we have presented refinements of S-money schemes
introduced in
Refs. \cite{kent2019s,Kentquantumtokens,Kentfuturepositioncommitment,Kentmethodandsystem}. 
These refinements make the schemes considerably more flexible and
practical.
They allow the initial commitment phase, which involves quantum
communications
and measurements and potentially may be relatively
lengthy, to take place long in advance.
This allows the user to choose her presentation
point at potentially any point in the intersection of the causal pasts
of all the presentation points, limited only by the need to 
complete the commitment by relatively short classical communications.   
It also allows the user to make a series of decisions, which may
be independent, that collectively determine the presentation point. 
This flexibility is valuable
in applications where speed is critical. 
Further, it allows S-money tokens to be simply and efficiently
transferred between users. 
The refinements retain the properties of unforgeability and 
future user privacy, both for the original user and for users
to whom the token is transferred.  

\begin{acknowledgments}
  The authors acknowledge financial support from the UK Quantum
  Communications Hub grant no. EP/M013472/1. AK is partially supported
  by Perimeter Institute for Theoretical Physics. Research at
  Perimeter Institute is supported by the Government of Canada through
  Industry Canada and by the Province of Ontario through the Ministry
  of Research and Innovation.
\end{acknowledgments}

\bibliography{relmoneyshort}

\end{document}